\documentclass[openacc]{rsproca_new}
\usepackage{tcolorbox}
\tcbuselibrary{skins, breakable}
\tcbset{
  highlightbox/.style={
    enhanced,
    colback=blue!5,
    colframe=blue!40!black,
    boxrule=0.8pt,
    arc=2mm,
    fonttitle=\bfseries,
    breakable,
    left=1mm,
    right=1mm,
    top=1mm,
    bottom=1mm
  }
}

\titlehead{Research}

\begin{document}

\title{Reproducible workflow for online AI in digital health}

\author{
Susobhan Ghosh$^{1}$*, Bhanu T. Gullapalli$^{1}$*, \\
Daiqi Gao$^{2}$, Asim Gazi$^{1}$, Anna Trella$^{1}$, \\
Ziping Xu$^{2}$, Kelly Zhang$^{3}$, \\
Susan A. Murphy$^{1,2}$
}

\address{$^{1}$Department of Computer Science, Harvard University, Cambridge, MA, USA, 02135\\
$^{2}$Department of Statistics, Harvard University, Cambridge, MA, USA, 02135\\
$^{3}$Department of Mathematics, Imperial College London, London, UK\\
$^{*}$Equal contribution}

\subject{Digital health, reproducibility, AI}

\keywords{Digital health, reproducibility, AI}

\corres{Susobhan Ghosh\\
\email{susobhan\_ghosh@g.harvard.edu}}

\begin{fmtext}
\begin{abstract}
Online artificial intelligence (AI) algorithms are an important component of digital health interventions. These online algorithms are designed to continually learn and improve their performance as streaming data is collected on individuals. Deploying online AI presents a key challenge: balancing adaptability of online AI with reproducibility. Online AI in digital interventions is a rapidly evolving area, driven by advances in algorithms, sensors, software, and devices. Digital health intervention development and deployment is a continuous process, where implementation—including the AI decision-making algorithm—is interspersed with cycles of re-development and optimization. Each deployment informs the next, making iterative deployment a defining characteristic of this field.   This iterative nature underscores the importance of reproducibility: data collected across deployments must be accurately stored to have scientific utility, algorithm behavior must be auditable, and results must be comparable over time to facilitate scientific discovery and trustworthy refinement. This paper proposes a reproducible scientific workflow for developing, deploying, and analyzing online AI decision-making algorithms in digital health interventions. Grounded in practical experience from multiple real-world deployments, this workflow addresses key challenges to reproducibility across all phases of the online AI algorithm development life-cycle.
\end{abstract}


\rsbreak


\section{Introduction}
\label{sec:intro}

 Digital health interventions -- or simply \emph{digital interventions} -- present an opportunity to address  substantial health challenges.
Leveraging ubiquitous technologies such as smartphones and wearable devices, these interventions are delivered directly to individuals as they go about their daily life. 
For example, individuals may receive reminders to take prescribed medication, prompts encouraging physical activity after long periods of inactivity, or supportive messages during periods of elevated stress. These actions—delivered through mobile notifications, SMS, or app-based interfaces—aim to promote healthy behaviors in their natural environment.
Digital interventions are now used across a wide range of health contexts, including depression \cite{himle2022digital,gunning2021digital}, medication adherence \cite{killian2019learning}, substance use \cite{ghosh2024miwaves,nesvaag2018feasibility,garland2023zoom, Gazi2022}, physical health \cite{liao2020personalized,stieger2023effects}, cardiovascular disease \cite{widmer2017digital}, cancer recovery \cite{roberts2017digital}, and many more.
Because the field of digital interventions
is an evolving area, driven by rapid advances in software, sensors  and wearable devices, the field  is inherently dynamic, driven by a process of \emph{continuous improvement} — that involves designing, testing, refining, and re-deploying interventions.  Data collected during each deployment plays a dual role: (1) it supports scientific discovery by enabling the evaluation of intervention efficacy and the analysis of individual behaviors,
and (2) it informs the (re)development of the intervention and any associated algorithm(s). Here, we use the term \emph{deployment} to refer to the
release and operation of a digital intervention in a real-world environment. This process of continuous improvement—driven by successive deployments and refinements is fundamental to how digital intervention research and practice evolve, enabling interventions to respond to not only  changes in operating systems, sensors and wearable devices, but also real-world societal changes and   individual diversity .

Challenges in this continuous improvement process are many.
Models or insights developed in one deployment may fail to transfer effectively to subsequent ones, due to changes in population characteristics, social dynamics, technological platforms, intervention delivery mechanisms and environmental contexts. Second, even within a single deployment, individuals differ widely in their needs, preferences and circumstances 
affecting their engagement with and response to interventions. To address these challenges, digital interventions have begun utilizing \emph{online} artificial intelligence (AI) decision-making algorithms\cite{el2019end,hwang2023personalization ,cancela2021digital}. Here,   the term \emph{online}  refer to AI decision-making algorithms that leverage streaming within and between individual data—such as passive sensing and self-reports— to learn parameters in a model for health outcomes and update estimates of these parameters repeatedly as new data becomes available during  deployment, in contrast to \emph{offline} algorithms, that learn model parameters on historical data which then remain fixed throughout deployment. 
Online AI decision making algorithms  make use of this online learning to make decisions regarding which and when to deliver interventions during deployment. 

\begin{figure}[!t]
    \centering
    \includegraphics[width=0.7\linewidth]{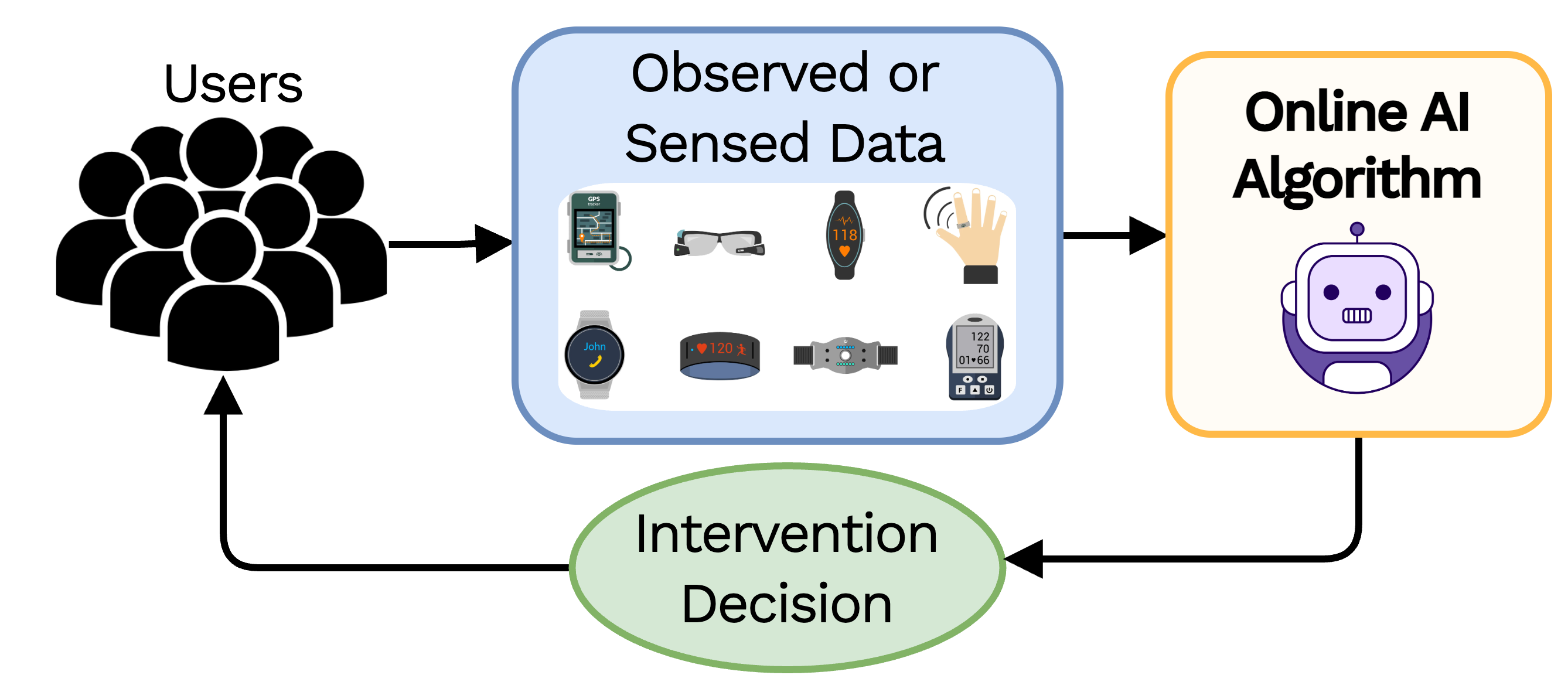}
\caption{Adaptive feedback loop of an online AI decision-making algorithm. Data collected during deployment informs the algorithm’s intervention decisions—such as sending a prompt, a supportive message, or a behavioral recommendation—which in turn influence the individual's future behavior and the data subsequently collected.}
    \label{fig:feedback_loop}
\end{figure}

Unlike traditional data collection settings, digital interventions employing online AI decision-making algorithms operate within an adaptive feedback loop: past data informs the algorithm’s current intervention decisions—such as sending a prompt or a message—which in turn influence individual biobehavioral outcomes  and thus the data collected in the future (see Fig \ref{fig:feedback_loop}) \cite{zhang2024replicable}. This dynamic interdependence introduces tightly coupled system behaviors, where seemingly minor changes can cascade into substantial shifts in data distributions, individual engagement, and intervention outcomes. As a result, the need for reproducibility becomes especially acute in such systems.

The continuous improvement of digital intervention deployments relies on the ability to learn from past deployments -- using collected data and observed outcomes to refine interventions and algorithms. 
For this process to support cumulative scientific progress and continuous data-driven improvements, the intervention system must produce data that is \emph{reproducible}. This paper follows the National Academies of Science, Engineering, and Medicine Consensus Report definition of reproducibility as the ability to obtain consistent results using the same input data, computational steps, methods, and conditions of analysis \cite{national2019reproducibility}. Critically, reproducibility is a prerequisite for \emph{replicability}, which is the ability to achieve consistent results across deployments aimed at answering the same scientific question but using independent data and potentially in a different implementation setting. While the National Academies definition provides a broad
foundation, reproducibility in online AI digital interventions is multifaceted. This paper distinguishes three complementary forms of reproducibility in online AI for digital health: \emph{design reproducibility}, \emph{deployment reproducibility}, and \emph{inference reproducibility}.
\begin{itemize}
    \item {\emph{Design reproducibility} : ability to rerun algorithm development and evaluation under identical conditions—for instance, replicating results in a digital twin testbed given the same input data, computational steps, and random seeds. It can be measured by verifying whether identical simulations and training pipelines yield identical outputs.}
    \item{\emph{Deployment reproducibility} : ability to reproduce system behavior during real-world operation—for instance, ensuring that identical inputs and model states yield identical decisions at runtime. It can be measured by verifying that decisions are fully traceable and auditable through logs, version control, and monitoring systems, including replaying logged states to confirm identical outputs or stress-testing monitoring systems to ensure that decision failures are consistently captured}
    \item{\emph{Inference reproducibility} : ability to obtain consistent scientific findings from deployment data, including analyses of algorithm performance and behavioral outcomes.}
\end{itemize}
These terms are introduced here for completeness, but the remainder of this paper focuses primarily on design and deployment reproducibility, as inference reproducibility with regards to inference based on analyses of existing data  has been addressed in some detail in the literature, for examples, see  \cite{yu2020veridical,yu2024veridical}. Reproducibility in the setting of online AI in digital health serves two complementary aims. 
First, it underpins \emph{scientific trustworthiness}, ensuring that results generated 
during deployment can be interpreted and verified. Second, it supports \emph{continuous 
improvement}, providing a stable and auditable foundation on which algorithms and 
interventions can be iteratively refined across deployments. These aims are aligned but not identical, and distinguishing between them clarifies how reproducibility contributes both to scientific validity and to system evolution.

This paper argues that reliable and stable infrastructure facilitates reproducibility. Reliability ensures that system behaviors remain consistent across contexts and time, thereby enabling scientific findings to be meaningfully interpreted and independently validated. Instability -- understood as the breakdown of reliability -- can be broadly categorized into two types: (1) engineering instability, stemming from issues such as inconsistent codebases, misaligned execution environments, or deployment mismatches; and (2) algorithmic instability, arising from learning procedures lacking practical convergence guarantees or abrupt changes to model logic. The adaptive feedback loop present in online AI decision-making algorithms can further exacerbate issues caused by both forms of instability, making the same recorded experimental inputs yield different and often unexpected outcomes at different times — ultimately challenging reproducibility and the iterative development cycle of online AI in digital interventions. Note that instability is not synonymous with uncertainty. In online settings, explicitly accounting for uncertainty—such as drift in the  distribution of health outcomes and variability in individual engagement and outcomes
—may yield less precise estimates, but conclusions that are ultimately more reproducible, trustworthy, and generalizable.

 To address these challenges, this paper proposes a reproducible workflow tailored for researchers and developers designing online AI digital interventions. The workflow provides structured guidance for building systems in which data and outcomes from each deployment can meaningfully contribute to ongoing scientific learning and system refinement. It is designed to help developers anticipate and manage the sources of instability identified above—ensuring that their systems not only function reliably in real time, but also support long-term, cumulative learning. The remainder of the paper is organized as follows: Section 2 provides background on the structure of online AI algorithms and the digital interventions in which they are deployed. Section 3 presents the proposed workflow, organized around key stages in the digital intervention system lifecycle. Section 4 offers practical recommendations drawn from real-world deployments, and Section 5 discusses broader implications, limitations, and future directions.

\section{Background}
This section describes the underlying components of online AI decision-making algorithms, along with examples of digital intervention involving AI algorithms that were deployed by the authors. The section begins by outlining two core components that are common across many such algorithms, followed by brief descriptions of real-world digital interventions that serve as running examples throughout the paper and ground the workflow in practical deployment contexts.



\subsection{Components of Online AI algorithms}
Online AI decision-making algorithms deployed in digital health interventions are typically composed of two interdependent components:
\begin{enumerate}
    \item \textbf{Online learning / Update procedures}: This component governs how the algorithm learns and adapts over time based on new data. It is responsible for updating the parameters of the algorithm's model using information collected during deployment. 
    The updates may occur continuously, at fixed intervals, or in response to specific triggers (e.g., user events or contextual changes), and are often subject to constraints on computation, memory, or communication.

    \item \textbf{Optimization / Decision-making}: This component determines, in real time, what intervention to deliver given the current context (e.g. data from sensors, sel-report data, etc.) and the algorithm's learned model parameters. It aims to optimize a predefined utility function -- such as maximizing individual engagement or behavior change -- while satisfying practical and ethical constraints on  intervention burden and safety. 
\end{enumerate}
\subsection{Reference Interventions}
\label{sec:past_int}
Throughout this paper, insights and examples are drawn from several real-world digital health interventions that the authors have been involved in. Each of these interventions deployed an online AI decision-making algorithm. These interventions will provide context for the workflow and recommendations discussed in later sections. Examples  were specifically selected to illustrate both the successes and failures of the authors in designing and deploying an online AI algorithm as part of a digital intervention evaluated in  a clinical trial: HeartSteps (ClinicalTrials.gov ID: NCT03225521), Oralytics (NCT05624489), and MiWaves (NCT05824754). In each example below, the goal of the intervention, the specific decision task assigned to the online AI—implemented as a reinforcement learning (RL) agent—and the contextual factors used to inform that decision are highlighted.
\paragraph{HeartSteps}~\ The HeartSteps digital intervention \cite{klasnja2015microrandomized,klasnja2019efficacy,dempsey2015randomised,liao2016sample} is designed to promote physical activity by delivering contextually tailored activity suggestions through smartphones and wearables. HeartSteps integrates a commercially available fitness tracker with a mobile application to sense individuals' behavior and context. HeartSteps uses an online RL \cite{liao2020personalized}  to decide, five times daily, whether or not to deliver an activity suggestion based on individuals' current context including prior intervention dose and physical activity.

\paragraph{Oralytics}~\ The Oralytics digital intervention \cite{nahum2024optimizing} is designed to improve  oral self-care behaviors in individuals at risk of dental disease. Oralytics integrates a commercially available electric toothbrush with Bluetooth connectivity and a mobile application to sense individuals' brushing behavior. An online RL algorithm \cite{trella2024oralytics,trella2025deployed} is utilized to determine, twice-daily, whether or not to deliver an intervention prompt  based  on recent brushing behavior, date and time, and recent engagement. 
\paragraph{MiWaves}~\ The MiWaves digital intervention \cite{coughlin2024mobile} is designed to help emerging at-risk adults reduce their cannabis use. MiWaves uses a mobile application to collect individuals' self-monitoring information.
MiWaves uses an online RL algorithm \cite{ghosh2024miwaves,ghosh2024rebandit} to determine, twice daily, whether or not to deliver an intervention prompt containing  behavior change strategies based on recent cannabis use, time of day and recent engagement in the behavior change strategies. 
\section{Workflow for Online AI in Digital Health}
\label{sec:workflow}



This section introduces an end-to-end workflow for developing, testing, deploying, and analyzing online AI-based decision-making algorithms in digital health, with reproducibility as a central design goal. The workflow is intended for researchers and practitioners building digital interventions that incorporate online AI, and who aim to ensure that the data, decisions, and outcomes generated during deployment can support scientific learning and continuous improvement. Illustrated in Figure~\ref{fig:workflow}, the workflow is structured into three key phases: (1) Designing the online AI decision-making algorithm, (2) Integration, assurance, and deployment in real-world environments, and (3) Post-deployment data analysis and inference.



\begin{figure}[!t]
    \centering
    \includegraphics[width=0.96\textwidth]{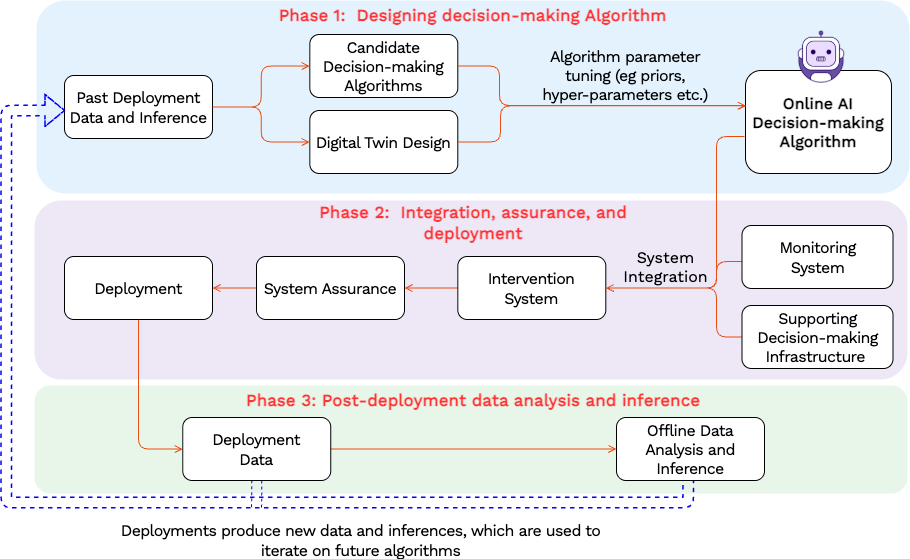} 
    \caption{ End-to-end reproducible workflow for online AI decision-making algorithms in digital health interventions. The workflow consists of three key phases: (1) algorithm design informed by past deployment data and digital twin(s), (2) system integration, assurance, and deployment with supporting infrastructure and monitoring systems, and (3) post-deployment data analysis and inference to evaluate performance and guide future algorithm and intervention iterations. 
    }
    \label{fig:workflow}
\end{figure}

\paragraph{Phase 1: Designing the online AI decision-making algorithm.}~\ 
Reproducibility considerations begin during the algorithm design phase, where \emph{past deployment data and inference} inform two key activities: (1) \emph{generating candidate decision-making algorithms}, and (2) \emph{evaluating the candidate decision-making algorithms in digital twin testbeds}.

The candidate algorithms
must account for real-world deployment constraints such as latency, bandwidth, and computational resources. These constraints influence not only how algorithms perform in deployment, but also what aspects of their behavior can be reliably reproduced. For instance, the interventions discussed in this paper (HeartSteps, Oralytics, and MiWaves) executed most of the online AI algorithm logic in the cloud, enabling greater control over computation, logging, and versioning. In contrast, interventions that rely on on-device computation and storage \cite{roy2022deep,maclean2013moodwings} must contend with stricter resource constraints and increased variability in execution environments -- such as limited processing power, memory, and storage capacity on wearables and smartphones. These limitations can directly impact reproducibility. For example when storage is limited, devices may be unable to store all intermediate computation outputs, or full decision logs. This may force system designers to incorporate lossy compression or selective storage of only the highest priority outputs for reproducibility,  making it impossible to perfectly reconstruct the algorithm's behavior post hoc. 

To ensure that candidate algorithms are robust and reproducible under such real-world conditions, their performance must be thoroughly evaluated prior to deployment. The \emph{Digital twin} plays a central role in this process.  It is constructed using historical deployment data to replicate key features of individual's behavior, contextual dynamics, and intervention responses \cite{Gazi2025}. By grounding the digital twin in \emph{empirical data distributions} -- such as engagement patterns, outcome trajectories, or contextual variable transitions -- researchers can construct a high-fidelity digital approximation of the deployment setting. Within this twin, one can instantiate multiple \emph{simulation environments} by systematically varying key parameters that represent real-world uncertainties. These include changes in the magnitude or variability of treatment effects, levels of individuals' heterogeneity, degrees of contextual drift, and patterns of data missingness or latency. Each combination of these factors defines a distinct environment, reflecting a scenario that domain experts and researchers might reasonably expect to encounter in deployment. These environments serve as a low-cost, low-stakes, and highly controllable sandbox, allowing algorithm designers to rapidly prototype, compare, and refine candidate algorithms without risking harm to individuals or incurring the overhead of live deployments.

Following the construction of the digital twin, the parameters of each candidate decision-making algorithm are tuned to optimize performance across multiple simulation environments in the testbed. This tuning process may include incorporating priors derived from past deployment data, performing hyper-parameter searches, etc. Once tuned, the candidate algorithms undergo rigorous evaluation across multiple simulation environments within the testbed. This evaluation enables the triage and selection of the most promising candidate(s) based on predefined criteria. In many cases, this includes multi-objective optimization, balancing factors like performance, fairness, and stability. The selected algorithm is then advanced to the next phase of the workflow: integration, testing and deployment. Taken together, these activities illustrate how Phase 1 contribute to \emph{design reproducibility}: by enabling 
algorithm development and evaluation to be repeated under identical conditions, such as rerunning digital twin 
simulations with the same input data, computational steps, and random seeds.

\paragraph{Phase 2: Integration, Assurance, and Deployment}~\ Once an online AI-based decision-making algorithm has been finalized, the focus shifts to embedding the algorithm within an \emph{intervention delivery system}. This integration phase is critical for enabling the online AI algorithm to interact with real-world individuals and data in real time.

The first step in this phase involves connecting the online AI decision-making algorithm to the broader \emph{supporting decision-making infrastructure}. This infrastructure typically includes three major components: data collection components (e.g., sensors and individual inputs), user interfaces, and the intervention backend. 
\emph{Data collection} components include both data from passive sensors -- from wearable devices and smartphones (e.g., GPS, accelerometer) -- and active inputs by individuals that capture self-reported mood, stress, cravings, etc. These diverse data streams may provide the contextual data necessary for the online AI-based decision-making algorithm to function effectively. \emph{User interfaces} may include applications through which the individual can interact with the intervention. Specifically in mobile health interventions, well-designed interfaces can be critical for individual engagement, adherence, and ultimately the success of the intervention \cite{hentati2021effect,wei2020design}. Finally, the \emph{intervention backend} coordinates all computational and logistical operations. This may include pre-processing raw sensor or self-report data into structured features, managing decision scheduling logic (e.g., determining when the algorithm's decision-making capabilities or update procedures should be triggered), and routing the assigned interventions to the individual's interface for delivery. Additionally, digital intervention systems may also incorporate a dedicated \emph{monitoring system}. A \emph{monitoring system} is a set of tools and processes used to track critical components of the digital intervention. Monitoring is essential in digital health applications, where individual safety and system reliability are paramount. A monitoring system typically logs decisions, system states, and errors, and may include alerting mechanisms to flag abnormal software behavior or intervention delivery failures \cite{national2018data,proschan2006statistical}.

Once the integration is complete, the intervention system undergoes extensive \emph{quality assurance} testing to ensure safe and reliable operation under realistic conditions. Testing may begin with automated simulations, where artificial inputs are streamed into the system to verify data flow, decision logic, and intervention delivery. This allows developers to identify and fix issues in a controlled environment before involving real individuals. The system may then progress to beta testing or pilot deployments, where a small group of individuals interacts with the intervention under supervised conditions. These tests help verify correct data processing, validate algorithmic responses to real-time contexts, and reveal potential issues such as message delays, sensor dropout, or latency. 
Following successful quality assurance testing, the intervention system incorporating the online AI decision-making algorithm is ready for real-world deployment.\\
Taken together, these activities contribute to \emph{deployment reproducibility}: ensuring that identical 
inputs and model states lead to the same intervention decisions, and that all system behaviors 
can be reconstructed and audited through comprehensive logging, monitoring, and version control.

\paragraph{Phase 3: Post-Deployment Data Analysis and Inference}~\  Subsequent sections do not address the reproducibility of phase 3 -- from deployment data pre-processing to offline analysis and inference -- as this topic has been extensively explored in prior work \cite{duncan2022veridicalflow,yu2020veridical,yu2024veridical} and falls outside the scope of this paper. However for completeness a brief overview of this phase is provided.  

After deployment is completed, the intervention system (incorporating the online AI decision-making algorithm) generates a stream of data—including contextual variables, algorithmic decisions, and observed outcomes—which is processed and analyzed offline. This analysis plays a critical role in interpreting system behavior and guiding future development. Specifically, this phase serves three primary purposes: (1) to evaluate algorithm performance and behavior, and (2) to guide future iterations of the algorithm and the intervention system , and (3) to support scientific discovery, including investigations into mechanisms of behavior change. 
Offline analysis leverages the deployment data—including contextual variables, algorithmic decisions, and observed outcomes—to answer a range of scientific and engineering questions. Depending on the evaluation goals, this analysis may involve statistical, causal inference, or machine learning methods to assess whether the algorithm achieved its intended outcomes \cite{ghosh2024did,trella2025deployed}, identify unintended consequences, or examine variation in effectiveness across subpopulations \cite{jones2020digital}. Techniques such as counterfactual modeling, heterogeneous treatment effect estimation, and mediation analysis are commonly used to support these objectives \cite{bendtsen2023mediators,domhardt2023mediators}. In addition, algorithmic stability is often evaluated to determine how sensitive the system’s decisions are to small changes in input data or modeling assumptions \cite{yu2013stability}, which is particularly important in high-dimensional or noisy settings.

Beyond evaluation, the collected data is used to refine future decision-making algorithms and intervention systems. This can involve retraining on new or more diverse datasets, updating model parameters to reflect the newly observed data distributions, or generating new candidate algorithms informed by recent deployment outcomes. Insights from offline analysis may also be used to construct more representative simulation environments, thereby improving the fidelity and robustness of subsequent simulation testbeds (Phase 1). 

Overall, this phase closes the loop in the online AI algorithm development workflow by feeding empirical evidence from real-world deployment back into the design phase for subsequent deployments. This iterative process enables data-driven refinement of digital health interventions, supporting both immediate enhancements and long-term system evolution. Although detailed discussion is beyond the scope of this paper, Phase 3 directly
involves \emph{inference reproducibility}, to ensure that analyses of deployment data can be
 interpreted and contribute to cumulative scientific progress.
 
\indent Finally, the above workflow can also be understood within the PCS framework for veridical data science 
\cite{yu2020veridical,yu2024veridical}, which emphasizes Predictability, Computability, and 
Stability as foundations for trustworthy analysis. In the digital intervention setting careful construction of digital twin environments, enactment of
pilot studies, and use of monitoring systems can  contribute to \emph{P (Predictability/Reality check)}, 
while practices such as version control, logging, and stability assessments can contribute to 
\emph{S (Stability)} by making decision-making reproducible and robust to perturbations in 
data and context. Considerations of infrastructure, compute resource management, and system integration 
can contribute to \emph{C (Computability)}. Here, recent work on 
simulation design under the MERITS framework \cite{elliott2024designing} provides additional practical 
guidance, particularly for the construction of digital twins that are modular, efficient, realistic, 
intuitive, transparent, and stable. Framing the above workflow through PCS highlights its alignment 
with broader principles of veridical data science. An illustrative case study of a sleep intervention with wearable sensors, showing how each phase of the workflow can be implemented in practice, is provided in the Supplementary material.

\section{Recommendations}
\label{sec:reco}

While the workflow in Section~\ref{sec:workflow} outlines how online AI decision-making algorithms can be developed and deployed in digital health settings, ensuring reproducibility in practice requires careful attention to system design, logging, and deployment procedures. This section offers concrete recommendations for improving reproducibility and reliability across each phase of the workflow. These recommendations are grounded in lessons from real-world deployments -- including HeartSteps, Oralytics, and MiWaves, described in Section ~\ref{sec:past_int} -- and address common failure modes and engineering challenges encountered in practice.

\subsection{Software Environments and Isolation}
\begin{tcolorbox}[highlightbox, title=Ensure workflow and system component isolation]
Ensure strict separation between development, testing, and deployment environments, and isolate system components (e.g., using containers such as Docker) to avoid interference and ensure stable operation.
\end{tcolorbox}
A \emph{software environment} refers to the complete configuration -- software libraries, operating system, dependencies, and runtime settings -- within which an AI algorithm or digital health intervention system operates. In the context of online AI in digital interventions, this includes both the environment in which algorithmic decisions are made and the broader system that supports data collection, processing, and intervention delivery. \emph{To ensure reliability in the intervention system, it is essential to maintain a clear separation between development, testing, and deployment software environments.} This separation ensures that modifications or instability during the development process do not accidentally propagate into deployment, and that testing reflects deployment conditions as closely as possible.
For instance, suppose developers are working on a new version of the decision-making algorithm in the development environment while an ongoing deployment is actively running. If the environments are not clearly separated—for example, if both versions share a codebase or infrastructure—a partially tested update from development environment might inadvertently be pushed into the current deployment. Imagine that this update introduces a bug that causes the algorithm to skip generating interventions under certain conditions (e.g., when recent sensor data is missing). This could lead to reduced intervention delivery, compromising both individual outcomes and the integrity of the collected data. Moreover, if the deployment logs do not reflect that a new version was unintentionally introduced, researchers analyzing the data post hoc may be unable to reconcile the algorithm’s behavior with the recorded system state, ultimately undermining reproducibility.

Beyond this vertical separation across workflow stages, \emph{a second form of separation -- component-level isolation within the deployment software environment -- is crucial for reliable functioning of online AI systems} during Phase 2: Integration, Testing, and Deployment. The digital intervention system typically comprises of multiple components running concurrently: the online AI decision-making algorithm, backend servers, databases, sensors, and user interfaces. Some of these components may be co-located on shared infrastructure (e.g., a cloud machine or mobile device), leading to potential interference through shared memory or processing resources. \emph{Isolation}, in this context, refers to the runtime separation of these components to prevent one from negatively impacting another's performance. Without isolation, failures or delays in one component can cascade into system-wide instability. For example, in the pilot study for MiWaves \cite{coughlin2024mobile,ghosh2024rebandit,ghosh2024miwaves}, both the algorithm and the intervention backend were deployed on the same machine without isolation. Because they shared memory and CPU resources, the system experienced memory management issues that eventually caused both components to crash -- negatively impacting the intervention delivery. While the developers involved in MiWaves systematically tracked these crashes with the help of a monitoring system (Section \ref{sec:fallback}), such instability can easily go unnoticed and untracked. If not recorded, these issues can silently alter the data being collected during deployment, impacting the scientific validity and reproducibility of data.
Ensuring such environment and component isolation is central to \emph{deployment reproducibility}, as it guarantees that system behavior can be replicated across environments, and any decisions made in deployment can be traced to a consistent and auditable execution context.
This type of instability can be effectively avoided by utilizing containerization and orchestration tools--such as containers managed with Docker, and orchestration frameworks like Kubernetes \cite{brewer2015kubernetes,merkel2014docker}. These tools are widely used to isolate critical system components, by allocating dedicated resources to each component, and enforcing strict software dependency versions independently.


\subsection{Autonomous Algorithm Design}
\begin{tcolorbox}[highlightbox, title=Design algorithms for autonomous operation]
Design algorithms to run autonomously with pre-specified update rules and 
stability guarantees, avoiding manual modifications during deployment to preserve scientific integrity.
\end{tcolorbox}
 In clinical trials for digital interventions, online AI decision-making algorithms are considered part of the intervention itself. As such, they are subject to constraints on modifiability -- particularly in clinical trials, where the intervention logic must be pre-specified in the study protocol. Once a trial has been launched, any changes to the algorithm's behavior or logic are typically prohibited to maintain scientific integrity \cite{trella2022designing}. To support this requirement, algorithms must be \emph{autonomous}, meaning they operate independently once deployed, without requiring manual intervention or modifications during the study period. In practice, this autonomy necessitates the use of computational procedures that provide \emph{stability guarantees}—ensuring that the algorithm's behavior remains controlled, interpretable, and robust to new data throughout deployment. Algorithms must be designed (in Phase 1) such that they are capable of learning online using well-defined update rules, without requiring mid-deployment calibration. This autonomy also directly supports \emph{deployment reproducibility}: by ensuring that the algorithm runs uninterrupted and unmodified throughout the trial, the system’s behavior remains consistent with what was pre-specified in the study protocol. Any observed outcomes can thus be reliably attributed to a fixed, well-documented process, rather than to ad hoc or undocumented adjustments.

For example, during the MiWaves pilot study, the algorithm designers utilized a simple Bayesian learning setup to derive closed-form posterior updates for model parameters, enabling stable online learning and avoiding convergence issues associated with more complex Bayesian inference techniques. Future work could use approximate inference methods like approximate Bayesian computation, for settings where closed-form posteriors are unavailable \cite{stephan2017stochastic,hoffman2013stochastic}. These approaches offer a promising direction for autonomous online learning with tractable approximation guarantees.


\subsection{Data Storage}
\begin{tcolorbox}[highlightbox, title=Log all data for full algorithm traceability]
Log all inputs, outputs, model states, update steps, and sources of randomness during deployment to enable exact post-hoc reconstruction of algorithm behavior.
\end{tcolorbox}

To ensure reproducibility in online AI decision-making algorithms for digital interventions, it is essential to systematically store all data used in the decision-making and update process.  This includes not only self-reported data or raw sensor streams, but also any processed features or summary statistics if those are the actual inputs to the algorithm. In addition, intermediate outputs -- such as decisions made, observed outcomes, model states, and update parameters should be logged.  Each data point, particularly those from smartphones or wearables, should be logged with its local device timestamp. Relying solely on backend timestamps can introduce inaccuracies due to communication delays, which may compromise reproducibility. Minor clock drift can typically be corrected post hoc if local timestamps are available.

When storing the entirety of reproducibility-critical raw data in its original form may not be feasible due to local storage constraints (e.g., sensor data sampled on the order of 10 Hz to 100 kHz), it is critical to thoughtfully employ the combination of lossless compression techniques, local or on-device storage (e.g., micro secure digital (SD) card), and asynchronous data synchronization with cloud storage. Without local storage, extended connectivity outages can lead to irreversible data loss, posing a serious risk to reproducibility. Compression can help reduce storage needs, but decompression procedures must be retained to allow accurate data reconstruction.

When raw data cannot be transmitted in real time due to bandwidth or power limitations, parts of the online AI algorithm -- such as decision logic or model updates -- may need to run on local devices capable of computation near the data source, often referred to as edge devices (e.g., a smartwatch or smartphone). These devices should locally store input summaries, decisions made, and model state, then transmit them asynchronously to the cloud when possible. To manage local storage, data should only be overwritten after successful cloud upload. In practice, uploads from such devices are best scheduled during charging periods or when the device is connected to Wi-Fi, to minimize power consumption and avoid transmission failures. This ensures that critical data are retained locally until reliable upload is possible, supporting reproducibility even in settings with intermittent connectivity.

One example that illustrates many of these challenges is the photoplethysmogram (PPG) signal, a common high-frequency input used in digital health interventions to estimate heart rate variability. The raw PPG signal reflects changes in peripheral blood volume, but it must be filtered, quality-checked, and segmented to extract inter-beat intervals. These pre-processing steps -- such as filter parameters, quality thresholds, and artifact removal procedures -- must be well-documented and consistently applied. Without this, variability in processing pipelines can produce inconsistent results that reflect software differences rather than underlying physiology.

Accurate timestamping is also critical. If local timestamps are not recorded at the device level, transmission delays can cause sampling intervals to appear irregular, compromising the integrity of the signal and violating reconstruction assumptions (e.g., Nyquist criteria). Inconsistent sampling or missing data due to connectivity loss may result in gaps or distortions that alter algorithm behavior. To avoid this, devices should retain data locally with high-resolution timestamps and only overwrite after confirming successful cloud upload. When connectivity is unavailable and the device cannot temporarily store data (i.e., lacks buffering support), permanent data loss can occur. In such cases, even recreating the same input context may yield divergent outcomes due to inconsistent data availability.

If storing raw data is infeasible, then the summaries or features used in decision-making must be logged, along with the code used to generate, compress, and decompress them. For probabilistic algorithms, additional elements -- such as random seeds, model parameters, and update timestamps -- must be recorded to enable exact reproduction of algorithm behavior and support post-deployment analysis.

Finally, it is essential to validate that algorithm updates are reproducible from stored data. Before deployment, researchers should confirm that the online learning pipeline produces consistent outputs from logged inputs. After deployment, these updates should be recreated offline to verify alignment. For example, in the Oralytics \cite{trella2024oralytics,trella2025deployed} and MiWaves \cite{coughlin2024mobile,ghosh2024miwaves} study, model parameters were reconstructed after each update and matched against deployment logs to confirm reproducibility throughout the trial. These validation procedures reinforce  
\emph{deployment reproducibility}, by ensuring that online decisions can be exactly replayed 
from logged states.

\subsection{Delayed and Missing Data for Decision Making}

\begin{tcolorbox}[highlightbox, title=Preserve real-time decision data]
Preserve the exact data version used in real-time decisions, including imputed values, and never overwrite with delayed ground-truth data to ensure reproducibility.
\end{tcolorbox}

Online AI decision-making algorithms deployed in digital interventions often utilize streaming data from sensors and individual inputs for decision-making. However, these data streams can be subjected to delays, dropouts or corruption during real-world deployments, which can impact algorithmic performance and compromise reproducibility if not properly accounted for. For example, in the Oralytics study, the decision-making algorithm used real-time toothbrush sensor data to deliver personalized interventions. However, delays in receiving the data occurred due to issues such as improper docking of the toothbrush and Bluetooth or Wi-Fi connectivity problems. Data was permanently lost in some cases because the toothbrush lacked onboard storage and could not transmit data to the cloud. Unfortunately, the toothbrush was a commercial product and thus, the team had no control over whether local storage was available onboard the toothbrush, as discussed in the prior subsection on data storage. Despite these challenges, the algorithm had to continue functioning to make decisions with delayed or missing inputs. These lapses underscore the need for robust strategies to handle delayed and missing data during deployment, both to maintain algorithm performance and to ensure reproducibility. 

When data are delayed or unavailable, it is critical to clearly document and archive the exact version of the data used for online learning or decision-making. If imputation is used to substitute missing inputs, the imputed values must be stored separately and preserved for reproducibility. These datasets should not be overwritten when the actual data becomes available later. 
Failure to do so can make it impossible to faithfully reconstruct how the algorithm operated in real time. 
For instance, in one deployment of HeartSteps \cite{ghosh2024did}, 
communication delays resulted in recent step-count information not being accessible to  the algorithm at the time the algorithm  needed to make decisions. In these scenarios, the online AI algorithm used imputed step-counts to make decisions. However, these imputed step count values were not logged, and later when the real step count information became available, it overwrote these imputed values, erasing the operational conditions under which real-time decisions had been made. As discussed in the limitations of a post-hoc analysis utilizing HeartSteps\cite{ghosh2024did} deployment data, the lack of imputation records made it impossible to reconstruct the decision parameters used by the algorithm at specific decision-times -- ultimately undermining the reproducibility of downstream scientific analyses. This example highlights why preserving the exact version of data used at decision-time 
is central to \emph{deployment reproducibility}. Without storing imputed or provisional 
inputs alongside later ground-truth values, it becomes impossible to replay the system’s 
behavior under actual operating conditions, breaking the audit trail needed for reliable 
analysis and interpretation.


\subsection{Algorithm Version Control}
\begin{tcolorbox}[highlightbox, title=Track algorithm logic with version control]
Use rigorous version control to document every logic or code change to the algorithm, and record the deployed version at all times for traceable decision reconstruction.
\end{tcolorbox}
During Phase 1, the online AI decision-making algorithm may undergo multiple revisions or modifications. Further changes can occur in Phase 2 (even during deployment) to address emerging issues, such as unforeseen technical constraints or bug fixes. These  changes differ significantly from the online parameter updates performed during algorithm optimization; rather, they reflect changes in the algorithm's core logic or structure. To ensure reproducibility, it is crucial to systematically track all such logic and code-level modifications using rigorous version control practices. It is equally important to document which specific software version of the algorithm was deployed at any given time. This typically involves maintaining a detailed, version-controlled repository -- using tools such as Git -- that records each change, including the specific code modifications, the rationale behind them, and corresponding timestamps. This applies also to code used for preprocessing input data from sensors and data collection tools, as well as data compression algorithms used to minimize storage requirements onboard edge devices and in the cloud. 

For instance, during the MiWaves pilot study, the algorithm developers identified a bug in the online AI algorithm that caused it to incorrectly determine the time of day for each decision point. The bug was subsequently fixed, and the corrected algorithm was thoroughly tested to ensure the issue no longer persisted. The corrected algorithm was assigned a new version number and was deployed while the study was still ongoing. This version change was appropriately recorded, ensuring that all subsequent decisions could be attributed to the correct algorithm version -- an essential step for maintaining reproducibility of the algorithm's decision-making. 
This example underscores the importance of having algorithmic version control to enable accurate reconstruction of decisions, verification of intervention outcomes, and overall transparency throughout the study. Rigorous version control therefore serves as a cornerstone of \emph{deployment reproducibility}, 
since it guarantees that every decision can be traced back to the exact algorithmic logic 
that produced it.


\subsection{Monitoring System and Fallback methods}
\label{sec:fallback}

\begin{tcolorbox}[highlightbox, title=Use monitoring and fallbacks for safe delivery]
Deploy a monitoring system to track algorithm behavior and support failure diagnosis, and define fallback methods to ensure consistent intervention delivery even when errors occur.
\end{tcolorbox}

While monitoring systems have long been used in clinical trials 
\cite{wiltsey2019frame,glasgow1999evaluating,mccreight2019using} -- to ensure intervention fidelity, assess participant engagement, and catch operational failures -- online AI decision-making algorithms introduce novel and under-addressed monitoring challenges. Here, \emph{intervention fidelity} refers to the extent to which an intervention is delivered as intended. In the context of online AI systems, this includes whether the algorithm generated decisions at the correct times, used the appropriate inputs, and behaved according to the deployed logic. The presence of an adaptive feedback loop significantly amplifies the potential consequences of system failures, data inconsistencies, or undetected bugs. Without a dedicated monitoring system for the online AI algorithm,   failures in intervention systems powered by online AI algorithms can propagate rapidly and silently, compromising both intervention fidelity, scientific validity, and reproducibility. These failures can  result in corrupted deployment data, missed or inappropriate interventions, and ultimately, invalid post-deployment analyses. In the worst-case scenario, such failures can lead to unintended participant harm, particularly if the AI algorithm is delivering biobehaviorally significant interventions. To address these challenges, recent frameworks have proposed systematic approaches to designing monitoring systems specifically for online AI decision-making algorithms in digital health \cite{trella2024effective}. 
By recording key metrics (e.g., model inputs, outputs, system states, errors, and exceptions) in a time-synchronized and queryable format, monitoring systems enable researchers to reconstruct algorithm behavior and validate whether the system functioned as intended throughout deployment. 

However, monitoring alone is insufficient. \emph{Fallback methods} are essential to ensuring that intervention delivery remains functional even when the online AI system encounters operational failures. These are predefined safety nets that activate in the presence of critical errors, missing data, or system crashes. Without them, system failures might be handled inconsistently or arbitrarily -- often through ad hoc human decisions that are undocumented and difficult to reproduce. This makes it nearly impossible to reproduce the algorithm's behavior or interpret the effects of interventions. By providing a consistent and well-documented response to failure conditions, fallback mechanisms help maintain the traceability and integrity of decision-making, even when the system is running in real-world conditions where things can go wrong.

For instance, in the Oralytics study \cite{trella2024oralytics,trella2025deployed}, a fallback intervention decision was executed if the online AI algorithm failed to retrieve the necessary data to form a decision. The fallback intervention strategy randomly selected among the available intervention options with equal probability. Although these interventions were not personalized using the online AI algorithm, this ensured that individuals would still receive an intervention decision, preserving both the reliability of the system and the completeness of collected data required for scientific discovery and iterative deployment. Together, monitoring systems and fallback methods are essential for 
\emph{deployment reproducibility}, by ensuring that system behavior and 
decision-making can be reconstructed even in the presence of failures.

\section{Discussion}
Reproducibility is not a purely academic concern. It is essential for both researchers and practitioners. For researchers, reproducibility enables re-analysis of past deployments, refinement of models, and valid causal inference — all of which depend on high-fidelity, reliable system execution. For industry stakeholders, it supports dependable product behavior, streamlined algorithm updates, and reduced operational risk. In digital health research, where interventions and associated algorithms are iteratively revised to accommodate evolving technologies and shifting real-world conditions, reproducibility and fidelity are indispensable.

This paper proposed a workflow aimed at addressing critical challenges related to the reproducibility and reliability of online AI decision-making algorithms in digital health interventions. These challenges arise both across deployments—where algorithms are repeatedly designed, tested, and refined—and within deployments, where algorithms must respond in real time to dynamic individual contexts, data irregularities, and system constraints. The workflow is structured around three key phases: (1) online AI algorithm design, (2) integration, assurance, and deployment, and (3) post-deployment analysis. Each of these phases maps onto a distinct reproducibility objective introduced in Section \ref{sec:intro}:
\emph{design reproducibility} in Phase 1, \emph{deployment reproducibility} in Phase 2,
and \emph{inference reproducibility} in Phase 3. This framing highlights how our workflow
operationalizes reproducibility across the full lifecycle of an online AI system for digital-interventions. Section~\ref{sec:workflow} introduces these phases and Section~\ref{sec:reco} offers concrete, field-tested recommendations based on real-world deployments of HeartSteps, Oralytics, and MiWaves. These include isolating software environments and system components, designing autonomous algorithms with stability guarantees, logging inputs and outputs for traceability, preserving real-time decision context despite data latency or dropouts, implementing version control for algorithm logic, and deploying monitoring systems and fallback methods. Together, these recommendations translate the abstract goal of reproducibility into a set of operational practices that can be implemented across diverse deployment settings.

At the same time, achieving reproducibility in practice is inherently constrained by real-world limitations. In many cases, digital interventions rely on commercial systems—such as wearable sensors, smartphones, or cloud platforms—whose internal software or data pipelines are not transparent or controllable. These systems may update their firmware or processing logic without notice, making is more difficult to achieve reproducible results even when the same inputs are presented. For example, a commercial wearable may alter its step-count algorithm or apply new filters without informing developers, breaking assumptions of consistency across deployments. These constraints must be acknowledged as inherent limits on reproducibility, particularly in intervention systems built atop proprietary platforms. 


In short, reproducibility must be treated as a central design goal in online AI for digital health--pragmatically pursued, transparently documented, and continuously evaluated. While perfect data reproducibility may not always be achievable, systems should be engineered to support it to the greatest extent possible. Doing so is essential not only for scientific validity and individual safety, but also for enabling the kind of continual learning and iterative refinement that underpins adaptive digital health interventions. As the field continues to evolve, centering reproducibility in the design and deployment of online AI systems will be key to building digital health interventions that are not only effective, but trustworthy, sustainable, and scientifically robust.

\ack{Research reported in this paper was supported by NIH/NIDA P50DA054039, NIH/NIDCR UH3DE028723, NIH/NHLBI R01HL125440, NIH/NIBIB K99EB037411, NIH/NIBIB and OD P41EB028242, and 5P30AG073107-03 GY3 Pilots. The content is solely the responsibility
of the authors and does not necessarily represent the official views of the National
Institutes of Health. Asim Gazi is supported by Schmidt Science Fellows, in partnership with the Rhodes Trust. Susan Murphy holds concurrent appointments at Harvard
University and as an Amazon Scholar. This paper describes work performed at
Harvard University and is not associated with Amazon. }

\bibliographystyle{RS}
\bibliography{sample}

\end{document}